\documentstyle[psfig,bibnorm]{lamuphys}
\begin{document}

\def\v{\vec{v}}
\def\x{\vec{x}}
\def\u{\vec{u}}
\def\e{\vec{e}}
\def\ro{\rho}
\def\be{\begin{equation}}
\def\ee{\end{equation}}
\def\P{\partial}
\def\d{\hbox{d}}


\title{Collective Motion}

\author{Andr\'as Czir\'ok and Tam\'as Vicsek}


\institute{Dept of Biological Physics, E\"otv\"os University, 1117 Budapest, P\'azm\'any stny 1, Hungary}

\maketitle

\begin{abstract}

With the aim of understanding the emergence of collective motion from local
interactions  of organisms in a "noisy" environment, we study biologically
inspired, inherently non-equilibrium models consisting of self-propelled
particles.  In these models particles interact with their neighbors by turning
towards the local average direction of motion.  In the limit of vanishing
velocities this behavior results in a dynamics analogous to some Monte Carlo
realization of equilibrium ferromagnets. However, numerical simulations
indicate the existence of {\it new types of phase transitions} which are not
present in the corresponding ferromagnets. In particular, here we demonstrate
both numerically and analytically that even in certain {\it one dimensional}
self-propelled particle systems an {\it ordered phase} exists for finite noise
levels.

\end{abstract}

\section{Introduction}

The collective motion of organisms (flocking of birds, for example), is a
fascinating phenomenon capturing our eyes when we observe our natural
environment.  In addition to the aesthetic aspects, studies on collective
motion can have interesting applications as well: a better understanding of the
swimming patterns of large schools of fish can be useful in the context of
large scale fishing strategies, or modeling the motion of a crowd of people
can help urban designers. Here we address the question whether
there are some global, perhaps universal features of this type of behavior when
many organisms are involved and such parameters as the level of perturbations
or the mean distance between the organisms is changed.

Our interest is also motivated by the recent developments in areas related to
statistical physics.  During the last 15 years or so there has been an
increasing interest in the studies of far-from-equilibrium systems typical in
our natural and social environment.  Concepts originated from the physics of
phase transitions in equilibrium systems \cite{HES71} such as collective
behavior, scale invariance and renormalization have been shown to be useful in
the understanding of various non-equilibrium systems as well.  Simple
algorithmic models have been helpful in the extraction of the basic properties
of various far-from-equilibrium phenomena, like diffusion limited growth
\cite{DLA}, self-organized criticality \cite{SOC} or surface roughening
\cite{surface}.  Motion and related transport phenomena represent a further
characteristic aspect of non-equilibrium processes, including traffic models
\cite{traffic}, thermal ratchets \cite{MM} or driven granular materials
\cite{gran}.

Self-propulsion is an essential feature of most living systems.  In addition,
the motion of the organisms is usually controlled by interactions with other
organisms in their neighborhood and randomness also plays an important role.
In Ref. \cite{VCBCS95} a simple model of self propelled particles (SPP) was
introduced capturing these features with a view toward modeling the collective
motion \cite{Reynolds87,motion} of large groups of organisms  such as schools
of fish, herds of quadrupeds, flocks of birds, or groups of migrating bacteria
\cite{AH91,bacbio,bacphys,CBCV,film}, correlated motion of ants \cite{millonas}
or pedestrians \cite{Helbing}.  Our original SPP model represents a statistical
physics-like approach to collective biological motion complementing models
which take into account much more details of the actual behavior of the
organism \cite{Reynolds87,jpn}, but, as a consequence, treat only a moderate
number of organisms and concentrate less on the large scale behavior.

In spite of the analogies with ferromagnetic models, the general behavior of SSP
systems are quite different from those observed in equilibrium models.  In
particular, in the case of equilibrium ferromagnets possessing continuous
rotational symmetry the ordered phase is destroyed at finite temperatures in
two dimensions \cite{MW66}.  However, in the 2d version of the SSP model phase
transitions can exist at finite noise levels (temperatures) as it was
demonstrated by simulations \cite{VCBCS95,CSV} and by a theory of flocking
developed by Toner and Tu \cite{TT} based on a continuum equation proposed by
them. Further studies showed that modeling collective motion leads to similar
interesting specific results in all of the relevant dimensions (from 1 to 3).
Therefore, after introducing the basic version of the model (in 2d) we discuss
the results for each dimension separately and then focus on the 1d case which
is better accessible for theoretical analysis.

\section{A generic model of two dimensional SPP system}

The model consists of particles moving on a plane with periodic
boundary condition. The particles are characterized by
their (off-lattice) location $\vec{x}_i$ and velocity $\vec{v}_i$
pointing in the direction $\vartheta_i$. To account for the self-propelled
nature of the particles  the magnitude of
the velocity is fixed to $v_0$. A simple local interaction  is
defined in the
model: at each time step a given particle assumes the average direction
of motion of the particles in its local neighborhood $S(i)$ with some
uncertainty, as described by
\begin{equation}
\vartheta_i (t+\Delta t) = \langle \vartheta(t) \rangle_{S(i)} + \xi,
\label{EOM}
\end{equation}
where the noise $\xi$ is a random variable with a uniform distribution
in the interval $[-\eta/2,\eta/2]$.
The locations of the particles are updated as
\begin{equation}
\vec{x}_i(t+\Delta t) = \vec{x}_i(t) + \vec{v}_i(t)\Delta t.
\label{update}
\end{equation}

The model defined by Eqs. (\ref{EOM}) and (\ref{update}) is a transport
related, non-equilibrium analog of the { ferromagnetic} models \cite{XY}.  The
analogy is as follows: the Hamiltonian tending to align the spins in the same
direction in the case of equilibrium ferromagnets is replaced by the rule of
aligning the direction of motion of particles, and  the amplitude of the random
perturbations can be considered proportional to the temperature for $\eta\ll
1$.  From a hydrodynamical point of view, in SPP systems the momentum of the
particles is not conserved. Thus, the flow field emerging in these models can
considerably differ from the usual behavior of fluids.

\section{Collective motion}

The model defined through Eqs.  (\ref{EOM}) and (\ref{update}) was studied by
performing large-scale Monte-Carlo simulations in Ref. \cite{CSV}. Due to the
simplicity of the model, only two control parameter should be distinguished:
the (average) density of particles $\varrho$ and the amplitude of the noise
$\eta$. 

For the statistical characterization of the configurations, a well-suited order
parameter is the magnitude of the average momentum of the system:
$\phi\equiv\left\vert \sum_j \vec{v}_j \right\vert/N$.  This measure of the net
flow is non-zero in the ordered phase, and vanishes (for an infinite system) in
the disordered phase.

Since the simulations were started from a disordered configuration,
$\phi(t=0)\approx 0$.  After some relaxation time a steady state emerges
indicated, e.g., by the convergence of the cumulative average $(1/\tau)
\int^\tau_0 \phi(t)dt$.  The stationary values of $\phi$ are plotted in Fig.~1a
vs $\eta$ for $\varrho = 2$ and various system sizes $L$.  For weak noise the
model displays long-range ordered motion (up to the actual system size $L$, see
Fig.2), that disappears in a continuous manner by increasing $\eta$.

As $L \rightarrow \infty$, the numerical results show the
presence of a kinetic phase transition described by
\begin{equation}
\phi(\eta)\sim \cases{
         \Bigl({\eta_c(\varrho) - \eta\over \eta_c(\varrho)}\Bigr)^\beta
                & for $\eta<\eta_c(\varrho)$ \cr
        0  & for $\eta>\eta_c(\varrho)$ \cr
    },
\label{scale}
\end{equation}
where $\eta_c(\varrho)$ is the critical noise amplitude that separates
the ordered and disordered phases and $\beta=0.42\pm0.03$, found to be
different from the the mean-field value $1/2$ (Fig~1b).

\begin{figure}
\centerline{
\hbox{
\psfig{figure=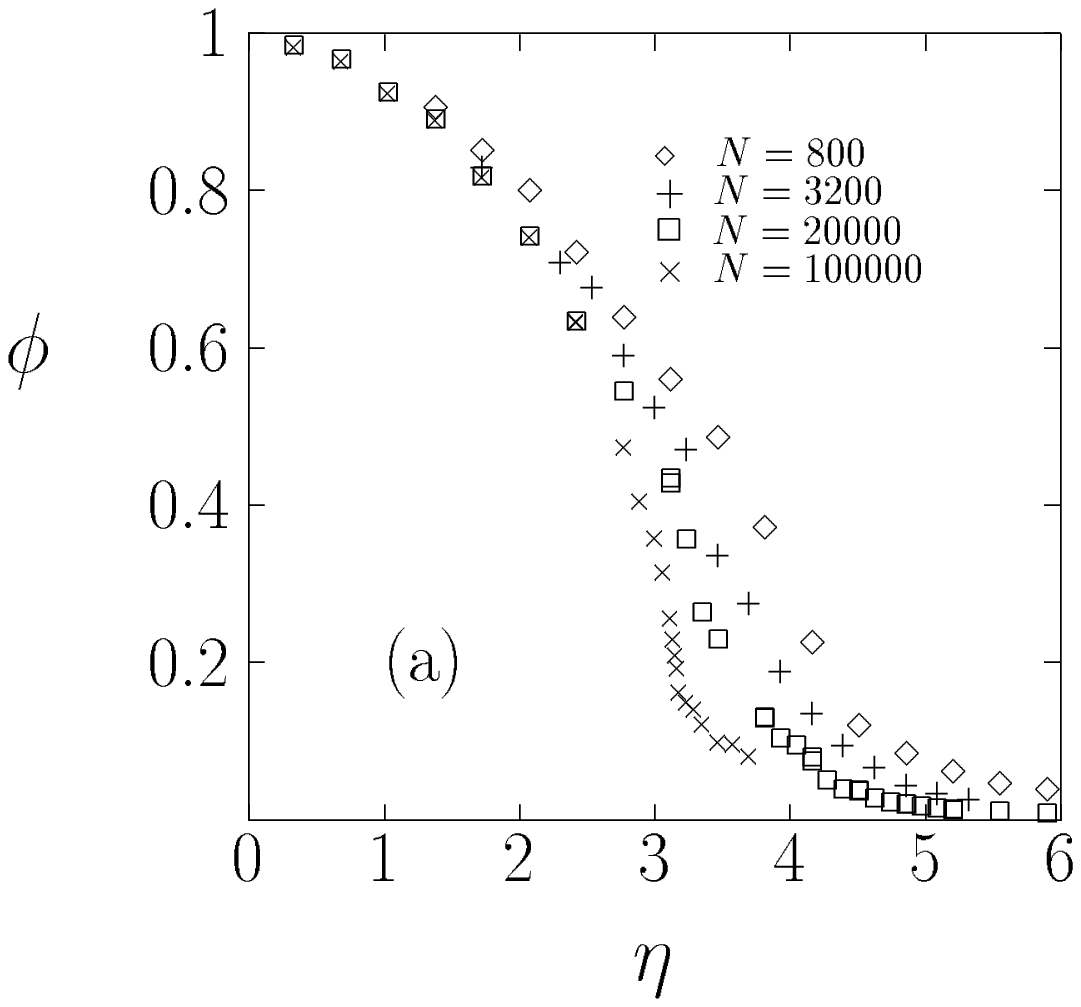,height=5.0truecm}
\psfig{figure=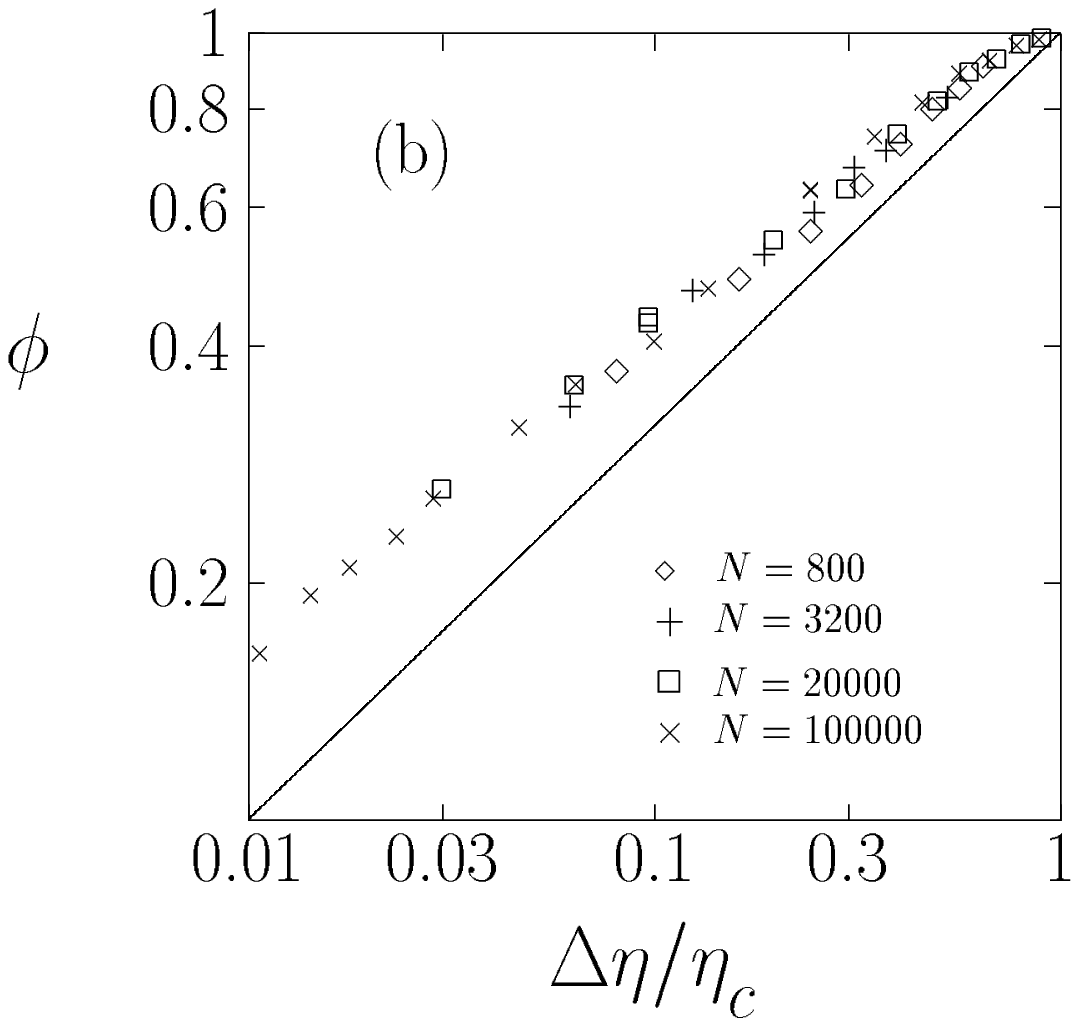,height=5.0truecm}
}}
\caption{(a) The average momentum of the 2D SPP model in the steady
state vs the noise amplitude $\eta$ for $\varrho = 2$ and four
different system sizes [($\diamond$) $N=800$, $L=20$; ($+$) $N=3200$,
$L=40$; ($\Box$) $N=20000$, $L=100$ and ($\times$) $N=10^5$, $L=223$].
(b) The order present at small
$\eta$ disappears in a continuous manner reminiscent of second order
phase transitions: $\phi\sim [(\eta_c(L)-\eta)/\eta_c(L)]^{\beta} \equiv
(\Delta\eta/\eta_c)^{\beta}$, with $\beta=0.42$, different from the
mean-field value $1/2$ (solid line).  }
\end{figure}

\begin{figure}
\centerline{
\hbox{
\psfig{figure=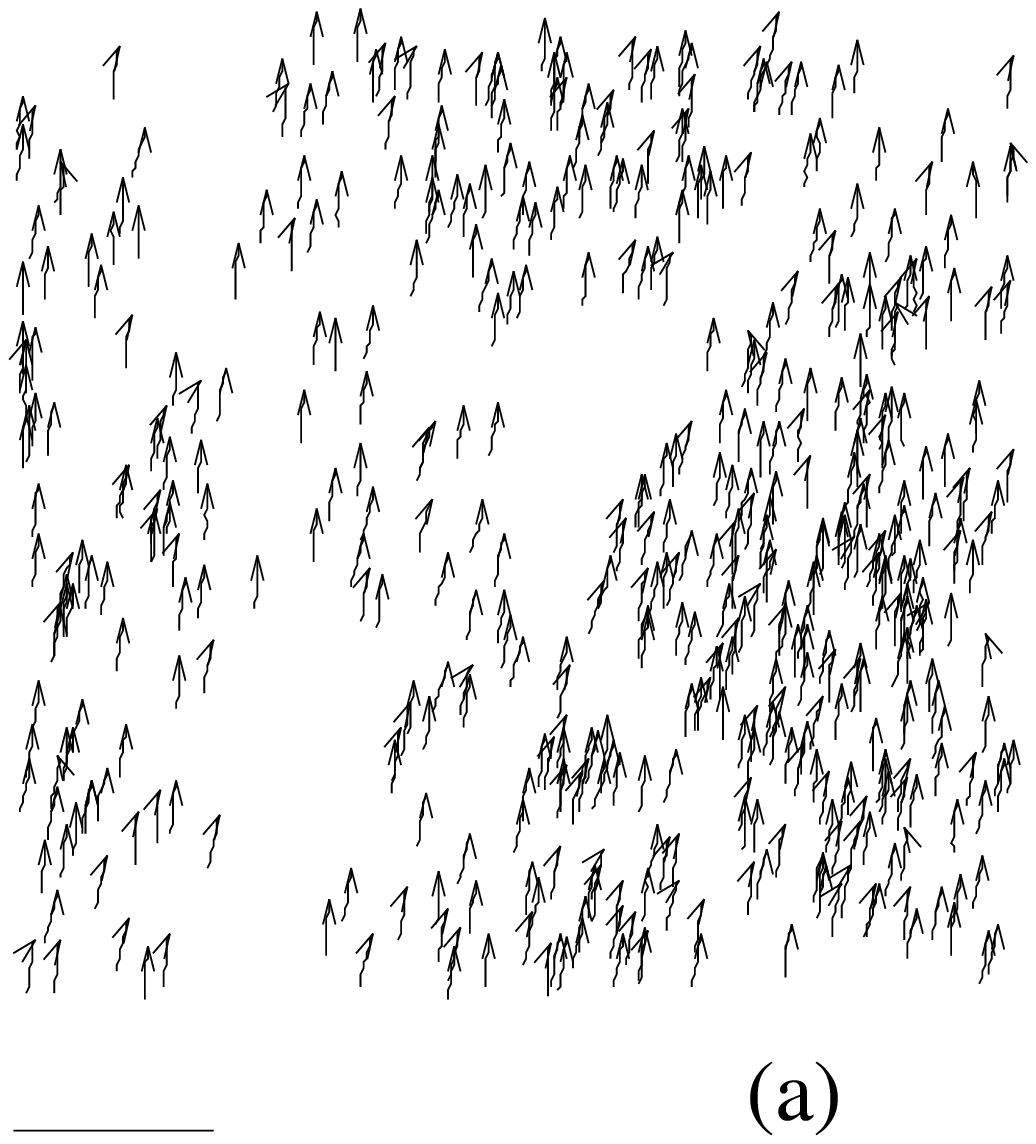,height=6.0truecm}
\psfig{figure=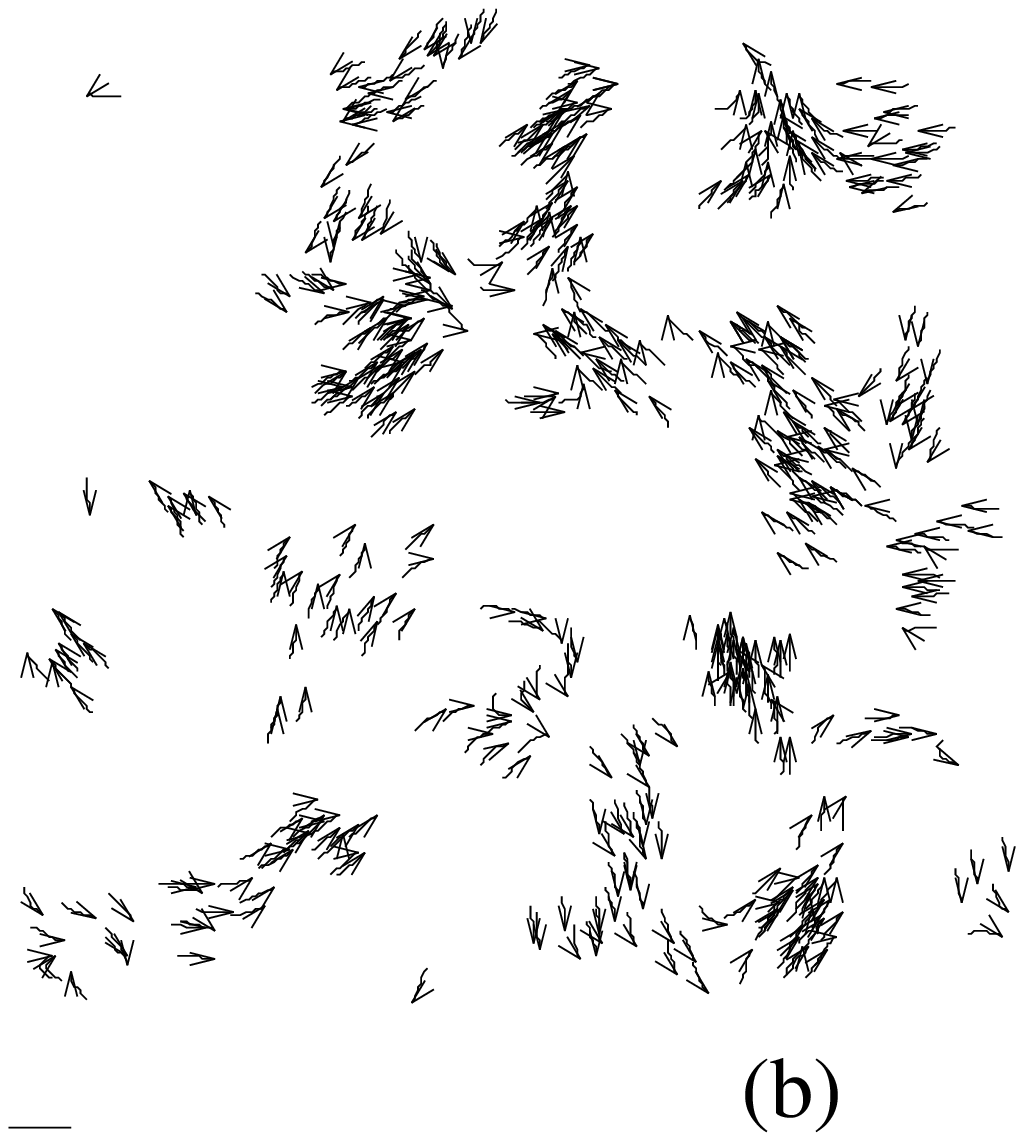,height=6.0truecm}
}}
\caption{The velocities of the SPPs are displayed for various values of
density and noise. The {actual velocity} of a particle is indicated by
a small arrow, while its trajectory for the last 20 time step is shown
by a short continuous curve. For comparison,
the range of the interaction is displayed as a bar.
(a) At high densities and small noise ($N=300$, $L=5$ and $\eta=0.1$) the
motion becomes ordered. (b) For small densities and noise
($N=300$, $L=25$ and $\eta=0.1$) the particles tend to form groups moving
coherently in random directions.
}
\end{figure}

In an analogy with equilibrium phase transitions, singular behavior of the
order parameter as a function of the density and critical scaling of the
fluctuations of $\phi$ was also observed. These results can be summarized as

\begin{equation}
\phi(\eta,\varrho) = \tilde{\phi}\bigl(\eta/\eta_c(\varrho)\bigr),
\label{v}
\end{equation}
where $\tilde{\phi}(x)\sim (1-x)^\beta$ for $x<1$, and $\tilde{\phi}(x) \approx
0 $ for $x>1$.  The critical line $\eta_c(\varrho)$ in the $\eta-\varrho$
parameter space was found to follow
\begin{equation}
\eta_c(\varrho)\sim\varrho^\kappa,
\label{kappa}
\end{equation}
with $\kappa = 0.45 \pm 0.05$.
Eq.(\ref{v}) also implies that the exponent
$\beta'$, defined as $\phi\sim(\varrho-\varrho_c)^{\beta'}$ for
$\varrho>\varrho_c$ \cite{VCBCS95}, must be equal to $\beta$.
The standard deviation ($\sigma$) of the order parameter behaved as
\begin{equation}
\sigma(\eta)\sim \vert \eta-\eta_c \vert^{-\gamma},
\end{equation}
with an exponent $\gamma$ close to $2$, which value is, again, different
from the mean-field result.

These findings indicate that the SPP systems can be quite well characterized
using the framework of classical critical phenomena, but also show surprising
features when compared to the analogous equilibrium systems.  The velocity
$v_0$ provides a control parameter which switches between the SPP behavior
($v_0>0$) and an $XY$ ferromagnet ($v_0=0$).  Indeed, for $v_0=0$
Kosterlitz-Thouless vortices \cite{KT} could be observed in the system, which
turned out to be unstable for any nonzero $v_0$ investigated in \cite{CSV}.

\section{Phase diagram for a 3d SPP system}

In two dimensions an effective  long range interaction can build up because the
migrating particles have a condiderably higher chance to get close to each
other and interact than in three dimensions (where, as is well known, random
trajectories do not overlap). The less interaction acts against ordering.  On
the other hand, in three dimensions even regular ferromagnets order.  Thus, it
is interesting to see how these two competing features change the behavior of
3d SPP systems.  The convenient generalization of Eq. (1) for the 3d case can
be the following \cite{CVV}:
\begin{equation}
\vec{v}_i(t+\Delta t)=v_0\hbox{ \bf N}(\hbox{ \bf N}(\langle
\vec{v}(t)\rangle_{S(i)}) + \vec\xi),\\
\label{EOM3D}
\end{equation}
where $\hbox{ \bf N}(\vec{u})=\vec{u}/\vert\vec{u}\vert$ and the noise
$\vec\xi$ is uniformly distributed in a sphere of radius $\eta$.

Generally, the behavior of the system was found \cite{CVV} to be similar to
that of described in the previous section.  The long-range ordered phase was
present for any $\varrho$, but for a fixed value of $\eta$, $\phi$ vanished
with decreasing $\varrho$.  To compare this behavior to the corresponding
diluted ferromagnet, $\phi(\eta,\varrho)$ was determined for $v_0=0$, when the
model reduces to an equilibrium system of randomly distributed "spins" with a
ferromagnetic-like interaction. Again, a major difference was found between the
SPP and the equilibrium models (Fig.~3): in the static case the system {\it
does not order} for densities below a critical value close to 1 which
corresponds to the percolation threshold of randomly distributed spheres in
3d.

\begin{figure}
\centerline{
\hbox{
\psfig{figure=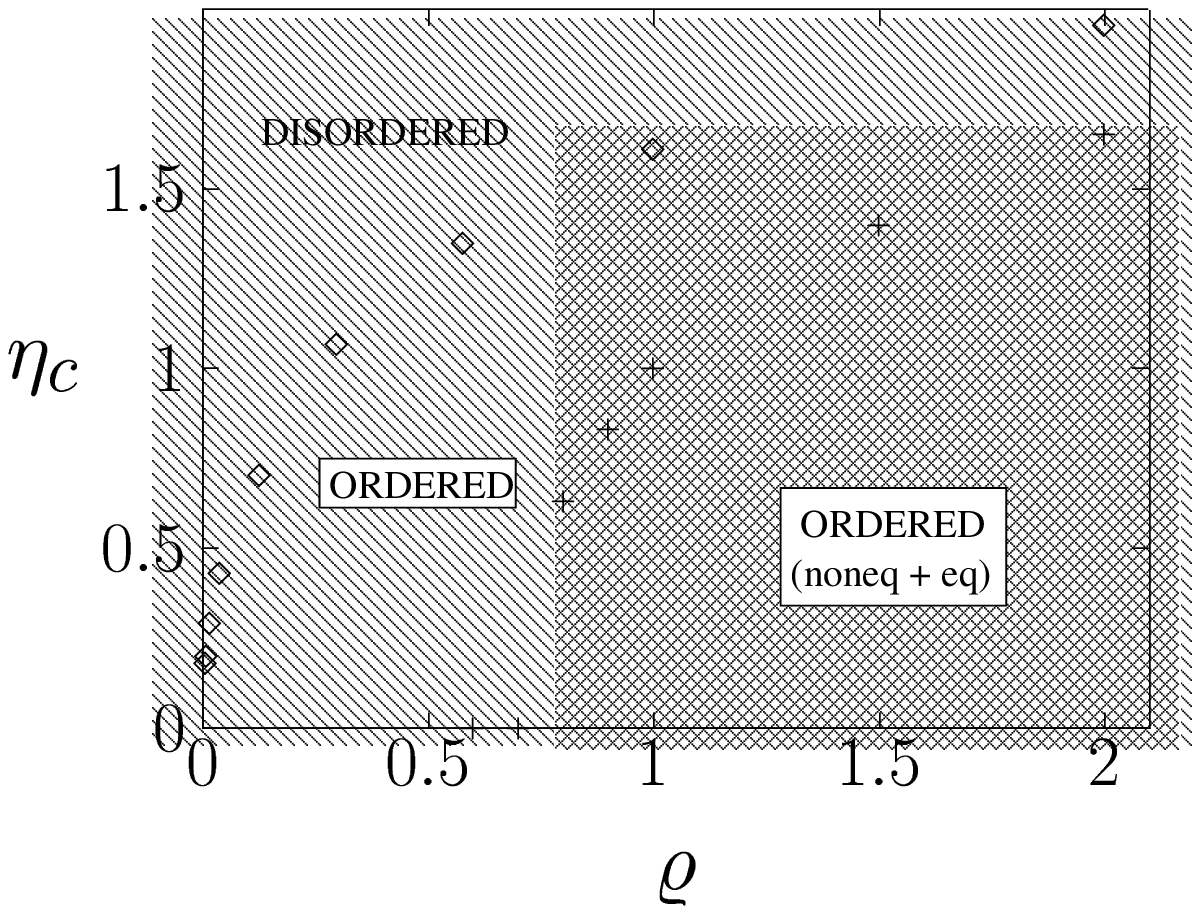,height=7.5truecm}
}}
\caption{
Phase diagram of the 3d SPP and the corresponding ferromagnetic system. 
The diamonds show our estimates for the critical noise for a given
density for the SPP model and the crosses show the same for the static
case. The SPP system becomes ordered in the whole region below the
curved line connecting the diamonds, while in the static case the
ordered region extends only down to the percolation threshold $\rho\simeq 1$.
}
\end{figure}

\section{Ordered motion for finite noise in 1d}

In order to study the 1d SPP model a few changes in the updating rules had to
be introduced. Since in 1d the particles cannot get around each other, some of
the interesting features of the dynamics are lost (and the original version
would become a trivial cellular automaton type model). However, if the algorithm
is modified to take into account the specific crowding effects typical for 1d
(the particles can slow down before changing direction  and  dense regions may
be built up of momentarily oppositely moving particles) the model becomes more
realistic.

Thus, in \cite{CBV} $N$ off-lattice particles along a line of length $L$ has
been considered. The particles are characterized by their coordinate $x_i$ and
dimensionless velocity $u_i$ updated as
\begin{eqnarray}
x_i(t+\Delta t) = x_i(t) + v_0 u_i(t)\Delta t, \\
u_i(t+\Delta t) = G\Bigl(\langle u(t) \rangle_{S(i)}\Bigr) + \xi_i.
\label{EOMD}
\end{eqnarray}
The local average velocity $\langle u \rangle_{S(i)}$ for the $i$th particle is
calculated over the particles located in the interval
$[x_i-\Delta,x_i+\Delta]$, where we fix $\Delta=1$.  The function $G$
incorporates both the propulsion and friction forces which set the velocity in
average to a prescribed value $v_0$: $G(u)>u$ for $u<1$ and $G(u)<u$ for $u>1$.
In the numerical simulations \cite{CBV} one of the simplest choices for $G$ was
implemented as
\begin{equation}
G(u)=\cases{
        (u+1)/2 & for $ u > 0$ \cr
        (u-1)/2 & for $ u < 0$, \cr
     }
\end{equation}
and random initial and periodic boundary conditions were applied.

\begin{figure}
\centerline{ \psfig{figure=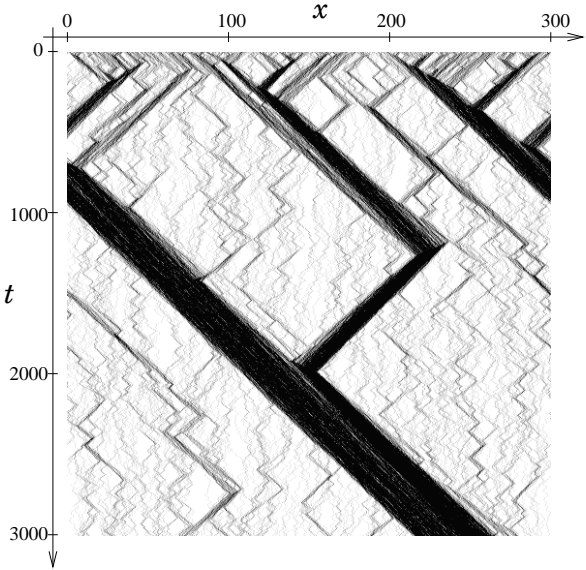,angle=0,height=7.5truecm} }
\caption{The first 3000 time steps of the 1d SPP model [$L=300$, $N=600$,
$\eta=2.0$ (a) and $\eta=4.0$ (b)]. The darker gray scale represents higher
particle density.}
\end{figure}

\begin{figure}
\centerline{
\hbox{
\psfig{figure=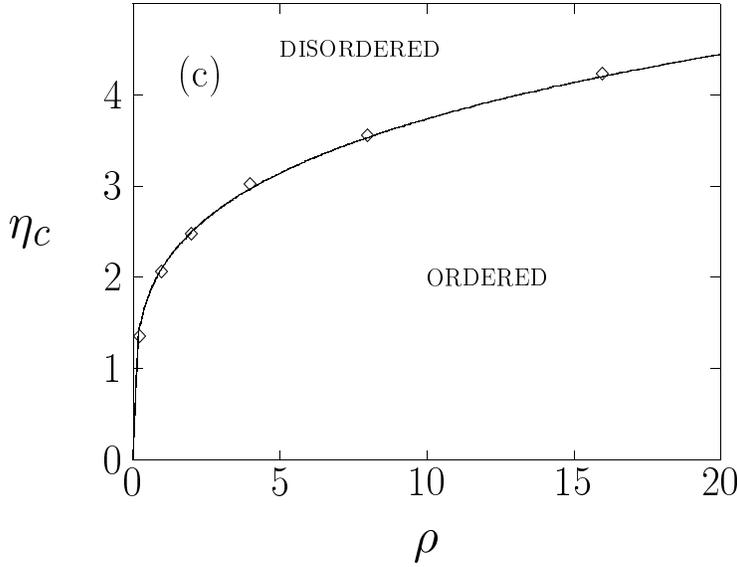,height=7.5truecm}
}}
\caption{
Phase diagram in the $\rho-\eta$ plane, the critical line follows
$\eta_c(\rho)\sim\rho^\kappa$. The solid curve represents a fit with
$\kappa=1/4$.  }
\end{figure}

In Fig.~4 we show the time evolution of the model for $\eta=2.0$. In a short
time the system reaches an ordered state, characterized by a spontaneous broken
symmetry and clustering of the particles. In contrast, for $\eta=4.0$ the
system would remain in a disordered state.  The $\ro-\eta$ phase diagram is
shown in Fig. 5. The critical line, $\eta_c(\varrho)$, follows (\ref{kappa})
with $\kappa\approx1/4$.

\section{Analytical studies of SPP systems}

\def\P{\partial}
\def\vr{\varrho}
\def\be{\begin{equation}}
\def\ee{\end{equation}}
\def\tu{\tilde{u}}
\def\hg{\hat{g}}
\def\hv{\hat{v}}
\def\hr{\hat{r}}
\def\hs{\hat{s}}
\def\ha{\hat{a}}
\def\hh{\hat{h}}
\def\vd{\vdots}

To understand the phase transitions observed in the models described
in the previous section, efforts has been made to set up a consistent
continuum theory in terms of $\v$ and $\rho$,
representing the coarse-grained velocity and density fields, respectively.
The first approach \cite{TT} has been made by J. Toner and Y. Tu
who investigated the following set of equations
\begin{eqnarray}
\P_t\v + (\v\nabla)\v &=& \alpha\v-\beta\vert\v\vert^2\v -\nabla P +
D_L\nabla(\nabla\v) + D_1\nabla^2\v + D_2(\v\nabla)^2\v+\xi\nonumber\\
\P_t\rho + \nabla(\rho\v)&=&0,
\label{TT1}
\end{eqnarray}
where the $\alpha,\beta>0$ terms make $\v$ have a nonzero magnitude,
$D_{L,1,2}$ are diffusion constants and $\xi$ is an uncorrelated
Gaussian random noise.
The pressure $P$ depends on the local density only, as given by the expansion
\be
P(\rho)=\sum_n\sigma_n(\rho-\rho_0)^n.
\ee
The non-intuitive terms in Eq.(\ref{TT1}) were generated by the
renormalization process.

Tu and Toner were able to treat the problem analytically and show the existence
of an ordered phase in 2d, and also extracted exponents characterizing
the density-density correlation function.  They showed that
the upper critical dimension for their model is 4, and the theory
does not allow an ordered phase in 1d.

However, as we have shown in the previous section, there exist SPP systems in
one dimension which exhibit an ordered phase for low noise level. Such systems
can not belong to the universality class investigated in \cite{TT}. This
finding motivated the construction of an other continuum model, which can be
derived from the master equation of the 1d SPP system \cite{CBV}:
\begin{eqnarray}
\partial_tU&=&f(U)+\mu^2\partial_x^2 U +
\alpha{(\partial_x U)(\partial_x \rho)\over\rho} + \zeta, \label{CEOM2}\\
\partial_t\rho &=& -v_0 \partial_x(\rho U) +
D\partial_x^2\rho \label{CEOM1},
\end{eqnarray}
where $U(x,t)$ is the coarse-grained dimensionless velocity field, the
self-propulsion term, $f(U)$, is an antisymmetric function with $f(U)>0$ for
$0<U<1$ and $f(U)<0$ for $U>1$.  The noise, $\zeta$, has zero mean and a
density-dependent standard deviation: $\overline{\zeta^2}=\sigma^2/\rho\tau^2$.
These equations are different both from the equilibrium field theories and from
the nonequilibrium system defined through Eqs(\ref{TT1}).  The main difference
comes from the nature of the coupling term $(\partial_x U)(\partial_x
\rho)/\rho$ which  can be {derived} from the microscopic alignment rule
(\ref{EOM}) \cite{CsCz}.  Note, that the noise also has different statistical
properties from the one considered in (\ref{TT1}).  For $\alpha=0$ the dynamics
of the velocity field $U$ is independent of $\rho$ and with an appropriate
choice of $f$ Eq.(\ref{CEOM2}) becomes equivalent to the $\Phi^4$ model
describing spin chains, where domains of opposite magnetization develop at
finite temperatures \cite{HES71}.

\section{Linear Stability Analysis}

To study the effect of the nonlinear coupling term $\alpha{(\partial_x
U)(\partial_x \rho)/\rho}$, we now investigate the development of the ordered
phase in the deterministic case ($\sigma=0$) when $c,D\ll1$ holds.  It can be
seen that stationary domain wall solutions exist for any $\alpha$. In
particular, let us denote by $\rho^*$ and $U^*$ the stationary solutions which
satisfy $\rho^*(\pm\infty)=0$, $U^*(\pm\infty)=\mp1$, $U^*(x)<0$ for $x>0$ and
$U^*(x)>0$ for $x<0$.  These functions are determined as
\be
\ln{\rho^*(x)\over\rho^*(0)} = {c\over D}\int_0^xU^*(x')dx'
\label{st1}
\ee
and
\be
\mu^2\P_x^2U^*=-f(U^*)-\alpha{c\over D}U^*\P_xU^*.
\ee
Although stationary solutions exist for any $\alpha$, they are
not always stable against long wavelength perturbations, as the
following linear stability analysis shows.

Let us assume that at $t=0$ we superimpose an $u(x,t=0)$ perturbation over the
$U^*(x)$ stationary solution. Since $c,D\ll1$ the dynamics of $\rho$ is slow,
thus $\rho(x,t)=\rho^*(x)$ is assumed.  The stationary solutions are metastable
in the sense that small perturbations can transform them into equivalent,
shifted solutions. Thus here by linear stability or instability we mean the
existence or inexistence of a stationary solution to which the system converges
during the response to a small perturbation. To handle the set of possible
metastable solutions we write the perturbation $u$ in the form of $\tu$ as
\be
U(x,t)=U^*(x)+u(x,t)=U^*(x-\xi(t))+\tu(x,t),
\label{pert}
\ee
i.e.,
\be
\tu(x,t)=u(x,t) + \xi(t)a(x),
\ee
where $a\leq0$ denotes $\P_xU^*$ and the position of the domain
wall, $\xi(t)$, is defined by the implicit
\be
U(\xi(t),t)=0
\label{imp}
\ee
equation. As $U^*(0)=0$, from (\ref{imp}) we have $\tu(\xi,t)=0$.
The usage of $\xi$ and $\tu$ is convenient,
since the stability of the stationary solution $U^*$ is equivalent
with the convergence of $\xi(t)$ as $t\rightarrow\infty$.

Substituting (\ref{pert}) into (\ref{CEOM1}) and taking into account
the stationarity of $U^*$
we get
\be
-a(x-\xi)\dot\xi + \P_t\tu(x) =(f'\circ U^*)(x-\xi)\tu(x) +
\mu^2\P_x^2\tu(x) +
\alpha\xi h(x-\xi),
\label{lin1a}
\ee
with $h=a\P_x^2\ln\rho^*\geq0$.  To simplify Eq.(\ref{lin1a} let us write
$(f'\circ U^*)(x)$ in the form of $g(x)-g_\infty$, where $g_\infty=
-lim_{x\rightarrow\pm\infty}f'(U^*(x))=-f'(\pm1)$. Furthermore, a moving frame
of reference $y=x-\xi$ and the new variable $v(y)=u(x)$ can be defined. With
these new notations Eq.(\ref{lin1a}) becomes
\be
-a(y)\dot\xi + \P_tv(y) = g(y)v(y) - g_\infty v(y) +
\mu^2\P_x^2v(y) + \alpha\xi h(y).
\label{lin2a}
\ee
The time development of $\xi$ is determined by the $u(\xi)=v(0)=0$ condition
yielding
\be
-a(0)\dot\xi = \mu^2\P_x^2v(0) + \alpha\xi h(0).
\label{lin2b}
\ee

By the Fourier-transformation of Eq. (\ref{lin2a}), treating the gv term as
a perturbation, for the time derivatives of $\xi(t)$ and the $n$-th Fourier
moments $\hv_n(t)\sim\P_x^nv(0,t)$ one can obtain \cite{linstab}
\be
{d\over dt}
\pmatrix{-\ha_0 &   &   &\cr
         -\ha_2 & 1 &   &\cr
	 -\ha_4 &   & 1 &\cr
	       \vd &   &   &\ddots \cr}
\pmatrix{\xi\cr \hv_2\cr \hv_4\cr \vdots} =
\pmatrix{\alpha \hh_0 & -\mu^2    &       	& \cr
	 \alpha \hh_2 & -g_\infty & -\mu^2	& \cr
	 \alpha \hh_4 &	& -g_\infty & -\mu^2 \cr
	 \vd		 & 	&	&\ddots \cr}
\pmatrix{\xi\cr \hv_2\cr \hv_4\cr \vdots}.
\label{det}
\ee

Expression (\ref{det}) can be further simplified using the relations
$\ha_m\ll\ha_n$ and $\hh_m\ll\hh_n$ for $m>n$ and the approximate
solutions for the $\lambda$ growth rate of the original $u$ perturbation
we found \cite{linstab}
\be
\lambda_0=\ha_0g_\infty<0
\ee
and $\lambda_+$,$\lambda_-$ satisfying
\be
\lambda_\pm^2 - b\lambda_\pm + q = 0,
\ee
where 
\begin{eqnarray}
b&=&\alpha\hh_0 - \mu^2\ha_2 + g_\infty \ha_0,\\
q&=&\alpha\ha_0(\hh_0g_\infty-\hh_2\mu^2).
\end{eqnarray}

If $\alpha=0$, then $\lambda_+=0$ and $\lambda_-=g_\infty \ha_0- \mu^2\ha_2<0$.
However, for certain $\alpha>0$ values $\lambda_+>0$ can hold  obtained as
either $q<0$ or $b>0$, i.e.,
\be
\alpha>\alpha_{c,1}=D{3g_\infty-2g(0)\over 2ca(0)}
\ee
or
\be
\alpha>\alpha_{c,2}=D{2g_\infty-g(0)\over a(0)(c-D)},
\ee
respectively. Thus for $\alpha>\alpha_c=min(\alpha_{c,1}, \alpha_{c,2})$
the stability of the domain wall solution disappears.

The instability of the domain wall solutions gives rise to the ordering
of the system as the following simplified picture shows.
A small domain of (left moving) particles moving opposite to the
surrounding (right moving) ones is {\it bound to interact} with more and
more right moving particles and, as a result, the domain wall assumes a
specific structure which is characterized by a buildup of the right
moving particles on its left side, while no more than the originally
present left moving particles remain on the right side of the domain
wall.  This process "transforms" non-local information (the size of the
corresponding domains) into a {\it local asymmetry of the particle
density} which, through the instability of the walls, results in a
leftward motion of the domain wall, and consequently, eliminates the
smaller domain.

This can be demonstrated schematically as

\begin{itemize}
\item[]{$>>>>>>>>>>> <<<<<<<< >>>>>>>>>>>>>$}

\item[]{$~~~~~~~~~~~~~~~~~~~~~~~~~A~~~~~~~~~~~~~~~~~B$}
\end{itemize}
where by $>(<)$ we denoted the right (left) moving particles.  In contrast to
the Ising model the $A$ and $B$ walls are very different and have
nonequivalent properties. In this situation the $B$ wall will break into a
$B_1$ and $B_2$, moving in opposite directions, $B_1$ moving to the left and
$B_2$ moving to the right, leaving the area $B_1-B_2$ behind, which is
depleted of particles.

\begin{itemize}
\item[]{$>>>>>>>>>>><<<<<~~~~~~~~~~~~>>>>>>>>>>$}
\item[]{$~~~~~~~~~~~~~~~~~~~~~~~~~A~~~~~~~B_1~~~~~~~~~B_2$}
\end{itemize}
At the $A$ boundary the two type of particles slow down, while,
due to the instability we showed to be present in the system, the wall itself
moves in a certain direction, most probably to the right.  Even in the other,
extremely rare case (i.e., when the $A$ wall moves to the left), an
elimination of the smaller domain ($A-B_1$) takes place since the velocity of
the domain wall $A$ is smaller than the velocity of the particles in the
"bulk" and at $B_1$ where the local average velocity is the same as the
preferred velocity of the particles.  Thus, the particles tend to accumulate
at the domain wall $A$, which again, through local interactions leads to the
elimination of the domain $A-B_1$.

It is easy to see that the $U=\pm1$ solutions are absolute stable against
small perturbations, thus it is plausible to assume that the system converges
into those solutions even for finite noise.

\section*{Acknowledgments}

We thank E. Ben-Jacob, H. E. Stanley and A.-L. Barabasi for useful
discussions. This work was supported by OTKA F019299 and FKFP 0203/1997.

\end{document}